\documentclass[letterpaper,twocolumn,english,aps,prl,final,superscriptaddress]{revtex4-1}
\usepackage[T1]{fontenc}
\usepackage[latin9]{inputenc}
\usepackage{xcolor}
\usepackage{pdfcolmk}
\usepackage{babel}
\usepackage{textcomp}
\usepackage{amsmath}
\usepackage{amssymb}
\usepackage{graphicx}
\PassOptionsToPackage{normalem}{ulem}
\usepackage{ulem}
\usepackage[unicode=true,
 bookmarks=true,bookmarksnumbered=false,bookmarksopen=false,
 breaklinks=true,pdfborder={0 0 0},pdfborderstyle={},backref=false,colorlinks=true]
 {hyperref}
\hypersetup{pdftitle={Origin of sawtooth domain walls in ferroelectrics},
 pdfauthor={D. Wang},
 pdfsubject={physics},
 pdfkeywords={domain wall, ferroelectrics}}

\makeatletter

\pdfpageheight\paperheight
\pdfpagewidth\paperwidth

\providecolor{lyxadded}{rgb}{0,0,1}
\providecolor{lyxdeleted}{rgb}{1,0,0}

\DeclareRobustCommand{\lyxsout}[1]{\ifx\\#1\else\sout{#1}\fi}

\usepackage{pslatex}
\allowdisplaybreaks
\usepackage{babel}
\usepackage{bm}
\usepackage{fixmath}

\usepackage{babel}

\usepackage{babel}

\makeatother

\begin{document}
\title{Origin of sawtooth domain walls in ferroelectrics}
\author{J. Zhang}
\affiliation{School of Microelectronics \& State Key Laboratory for Mechanical
Behavior of Materials, Xi'an Jiaotong University, Xi'an 710049, China}
\author{Y.-J. Wang}
\affiliation{Shenyang National Laboratory for Materials Science, Institute of Metal
Research, Chinese Academy of Sciences, Wenhua Road 72, 110016 Shenyang,
China}
\author{J. Liu}
\affiliation{State Key Laboratory for Mechanical Behavior of Materials, School
of Materials Science and Engineering, Xi'an Jiaotong University, Xi'an
710049, China}
\author{J. Xu}
\affiliation{School of Microelectronics \& State Key Laboratory for Mechanical
Behavior of Materials, Xi'an Jiaotong University, Xi'an 710049, China}
\author{D. Wang}
\email{dawei.wang@xjtu.edu.cn}

\affiliation{School of Microelectronics \& State Key Laboratory for Mechanical
Behavior of Materials, Xi'an Jiaotong University, Xi'an 710049, China}
\author{L. Wang}
\affiliation{Electronic Materials Research Laboratory, School of the Electronic
and Information Engineering, Xi'an Jiaotong University, Xi'an 710049,
China}
\author{X.-L. Ma}
\affiliation{Shenyang National Laboratory for Materials Science, Institute of Metal
Research, Chinese Academy of Sciences, Wenhua Road 72, 110016 Shenyang,
China}
\author{C.-L. Jia}
\affiliation{School of Microelectronics \& State Key Laboratory for Mechanical
Behavior of Materials, Xi'an Jiaotong University, Xi'an 710049, China}
\affiliation{\textsuperscript{}Ernst Ruska Center for Microscopy and Spectroscopy
with Electrons, Research Center Jülich, D-52425 Jülich, Germany}
\author{L. Bellaiche}
\affiliation{Department of Physics and Institute for Nanoscience and Engineering,
University of Arkansas Fayetteville, Arkansas 72701, USA}
\date{\today}
\begin{abstract}
Domains and domain walls are among the key factors that determine
the performance of ferroelectric materials. In recent years, a unique
type of domain walls, i.e., the sawtooth-shaped domain walls, has
been observed in BiFeO$_{3}$ and PbTiO$_{3}$. Here, we build a minimal
model to reveal the origin of these sawtooth-shaped domain walls.
Incorporating this model into Monte-Carlo simulations shows that (i)
the competition between the long-range Coulomb interaction (due to
bound charges) and short-range interaction (due to opposite dipoles)
is responsible for the formation of these peculiar domain walls and
(ii) their relative strength is critical in determining the periodicity
of these sawtooth-shaped domain walls. Necessary conditions to form
such domain walls are also discussed.
\end{abstract}
\maketitle
Domains, which are typical regions with aligned magnetic moments or
electric dipoles, can largely influence phase transitions and physical
properties of magnetic or ferroelectric materials. For ferroelectrics,
many attentions have been paid to investigate domains' characteristics
and properties \citep{Merz,w-2,w-3,w-4,w-5}. When changing from the
paraelectric to the ferroelectric phase, the symmetry of equivalent
dipole directions is broken, giving rise to regions with different
polarization directions while each region has a preferred polarization
direction. Ferroelectric domain walls have received extensive attention
due to various novel phenomena, including stable patterns on the nanometer
scale. Domains have been carefully analyzed to reveal the correlation
between the micro/nanoscale structure and the properties of the materials
\citep{w-6,w-7,w-8,w-9}, often through high resolution X-ray diffraction
technique \citep{w-14,w-15}. For instance, polarization switching
is a critical link between domains and material performance \citep{w-11,w-12,w-13,LB1,LB2,LB3}.
In bulk ferroelectrics, the domain structure, closely related to phase
structure, was thoroughly discussed along with domain size and morphology.
On an even smaller scale, polar nanoregions as a special type of domains
have also been discussed \citep{w-16,w-17,w-18,w-19,LB4,LB5}.

Recently, charged domain walls have attracted much investigation\citep{chargedm1,chargedm2}.
They are a special type of interface between two domains that is ultrathin
(usually on the nanoscale), where bound carries arise from the abrupt
change of polarization causing discontinuity on the interface. For
instance, the head-to-head polarization leads to a positive charge
while tail-to-tail configuration results in negative charge. Due to
the competition between the electrostatic energy (aligned dipoles
usually have smaller electrostatic energy) and the domain wall energy
(the extra energy necessary to have domains), domains can have very
different morphologies, such as rhombohedral, orthorhombic, and tetragonal
domains \citep{domain}. However, it was still quite surprising when
sawtooth-shaped 180$^{\circ}$ domain walls were observed in multiferroic
BiFeO$_{3}$ (BFO) (see Fig. 4(a) of Ref. \citep{Jia2015}), which
has a spontaneous polarization along the pseudocubic $\left\langle 111\right\rangle _{c}$
direction (that can be as large as 90-95 $\mu\textrm{C}/\textrm{cm}{}^{2}$
\citep{Chu2007}) and a high Curie temperature ($T_{C}=820\thinspace{^{\circ}}\textrm{C}$)
\citep{Wang_J_2003,BFO2,Wang2012}. Note that the BiFeO$_{3}$ sample
of Ref. \citep{Jia2015} was cut along $\left\langle \bar{1}\bar{1}0\right\rangle $
and $\left\langle 1\bar{1}0\right\rangle $ while extending 55 nm
vertically when high resolutiontransmission electron microscopy (HRTEM)
images were taken. More recently, Zou \emph{et al.} \citep{chinese_sawtooth}
also observed serrated 180$^{\circ}$ domain walls in PbTiO$_{3}$
(PTO) thin films prepared by pulsed laser deposition. This PTO thin
film was 100 nm thick and epitaxially grown on a (100)-oriented single
crystal SrTiO$_{3}$ substrate (see Figs. 2 and 4 of Ref. \citep{chinese_sawtooth}).
These observations indicate that sawtooth-shaped domain wall constitute
a general phenomenon in ferroelectrics, not limited to multiferroics
or magnetic materials \citep{Curland1970,Moon2018}. The bound charge
on domain walls can be quite large (for BFO, the bound charge is estimated
to be 1.64\,$\left|e_{0}\right|$, where $e_{0}$ is the electron
charge \citep{estimate_bound_change}), which can strongly affect
the conductivity of the material by attracting free charge carriers,
making them good candidates for domain wall electronics \citep{Seidel2009,N. Setter1,LB6}.
Recent research also show that negative capacitance is also closely
related to dipole patterns and domain structures \citep{N. Setter2,Lukyanchuk2019}.

While charged domain walls have been known for a long time \citep{Vul1973}
and different aspects had been investigated including their conductivity
\citep{Seidel2009,Sluka2013,chargedm2,Rojac2017,chargedm1}; their
influence on other dipoles and enhancement of material performance
\citep{charged-domain-walls,Li2013}; their dynamics of the charged
domain wall and interaction with electric field \citep{Shur2000,charge_domain2,Gureev2012,McQuaid2012,Esin2017},
we are not aware of theoretical work to explain the formation mechanism
of the sawtooth-shaped charged domain walls in ferroelectric materials.
In this work, we explore possible causes of this unique phenomenon
of sawtooth-shaped domain walls, finding that the long-range Coulomb\textbf{
}between bound charges and the short-range interaction between opposite
dipole pairs are adequate to reproduce such peculiar domain walls.

\begin{figure}[h]
\begin{centering}
\includegraphics[width=8cm]{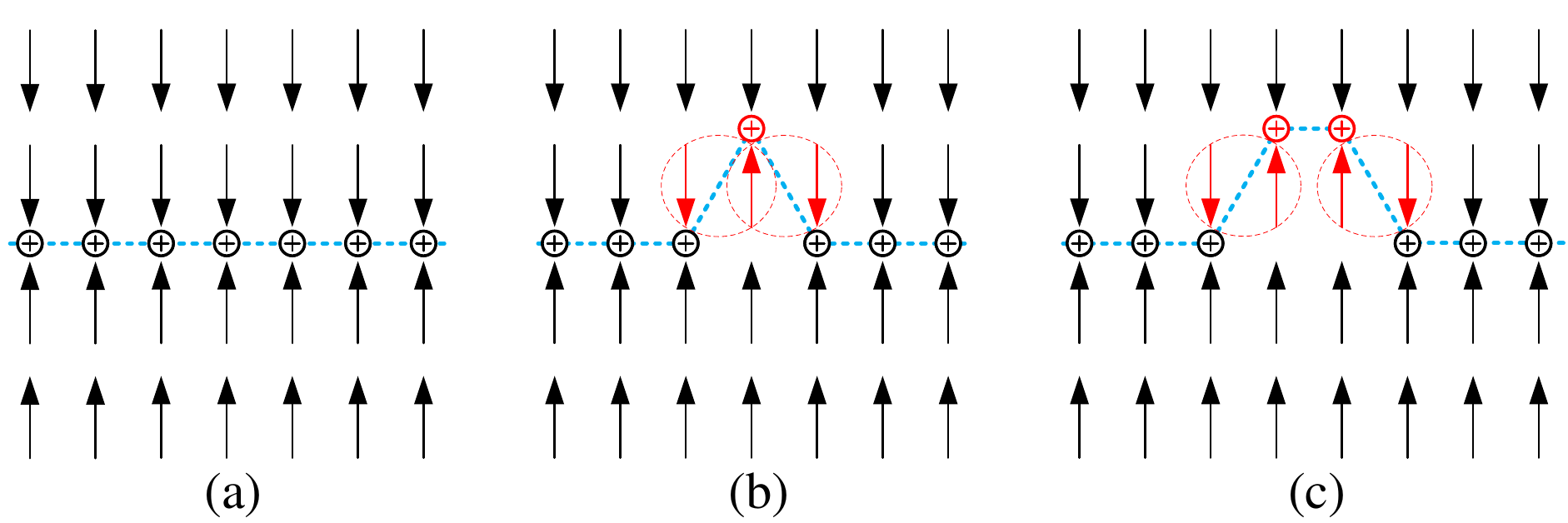}
\par\end{centering}
\caption{\label{fig:The-model-of-domain-wall}Schematic drawing of dipoles
and bound charges. (a) The black arrows represent dipoles, while the
blue line depicts the 180${^{\circ}}$ domain wall. The symbol ``{\footnotesize{}$\bigoplus$''}
between dipoles represents bound charges formed by head-to-head dipoles.
(b) When a dipole is reversed, the domain wall and the position of
one bound charge change accordingly, and two pairs of opposite electric
dipole pairs are generated at the left and right sides of the reversed
dipole. (c) Another configuration can also involve two pairs of opposite
dipoles.}
\end{figure}
 As a matter of fact, in order to understand the sawtooth domain walls,
we propose a minimal model with just short-range interaction between
opposite dipoles and long-range Coulomb interaction due to bound charges
arising from the head-to-head dipoles, and following a similar approach
as the effective Hamiltonian \citep{Zhong,key-3,Wang2013,key-5,Wang2014,Wang2016}
to simulate 2D and 3D ferroelectric materials. We assume that (i)
electric dipoles of opposite directions already exist in the system,
and (ii) a boundary exists between the two groups of opposite dipoles
(see Fig. \ref{fig:The-model-of-domain-wall}). As bound charges accumulate
on the boundary, their positions can be used as dynamic variables
in simulations while the number of bound charges is fixed, which determines
both the Coulomb energy and the short-range interaction as Fig. \ref{fig:The-model-of-domain-wall}
shows. Therefore, the total energy for the system is given by:

\begin{align}
E^{\textrm{tot}} & =E^{\textrm{cc}}\left(\left\{ \boldsymbol{r}_{i}\right\} \right)+E^{\textrm{short}}\left(\left\{ \boldsymbol{r}_{i}\right\} \right)\label{eq:total-energy}
\end{align}
where $\boldsymbol{r}_{i}$ is the position of the $i$th bound charge.
$E^{\textrm{short}}$ is the short-range energy when neighboring ions
have relative shifts \citep{Zhong}. For the 2D case shown in Fig.
\ref{fig:The-model-of-domain-wall}, the short-range interaction on
the domain wall can be expressed as $E^{\textrm{short}}=JN$, where
$J>0$ is the additional energy associated with opposite neighboring
dipoles and $N$ (depending on $\left\{ \boldsymbol{r}_{i}\right\} $)
is the number of opposite dipole pairs. $E^{\textrm{cc}}=\frac{1}{2}\sum_{i,j}Z^{2}/\varepsilon_{r}\left|\boldsymbol{r}_{i}-\boldsymbol{r}_{j}\right|$
is the long-range charge-charge Coulomb energy, where $Z$ is the
bound charge, $\varepsilon$ is the relative permittivity, and the
energy unit is Hartree. Since the sawtooth domain wall induces bound
charges and opposite dipole pairs, $E^{\textrm{tot}}$ can also be
regarded as the formation energy of the domain wall. For simplicity,
we use the energy of Fig. \ref{fig:The-model-of-domain-wall}(a) as
the reference energy $E_{0}$, implicitly subtracting $E_{0}$ from
$E^{\textrm{tot}}$ hereafter. It shall be emphasized that the proposed
energy terms constitute a minimal model that, as we will show, demonstrate
why sawtooth domain wall arises. More sophisticated phenomenological
model will be discussed in the Supplemental Material.

Using the total energy of Eq. (\ref{eq:total-energy}), Monte-Carlo
(MC) simulations are employed to find the equilibrium domain wall
morphology. During the simulation, the position of the bound charges
($\boldsymbol{r}_{i}$) are tracked and changed to minimize the free
energy. In each MC simulation at 300 K, we perform 320,000 sweeps
of all the $\boldsymbol{r}_{i}$. We will first show the simulation
results and then discuss how the parameters ($J$ and $Z$) can affect
the morphology.

\begin{figure}[h]
\begin{centering}
\includegraphics[width=8cm]{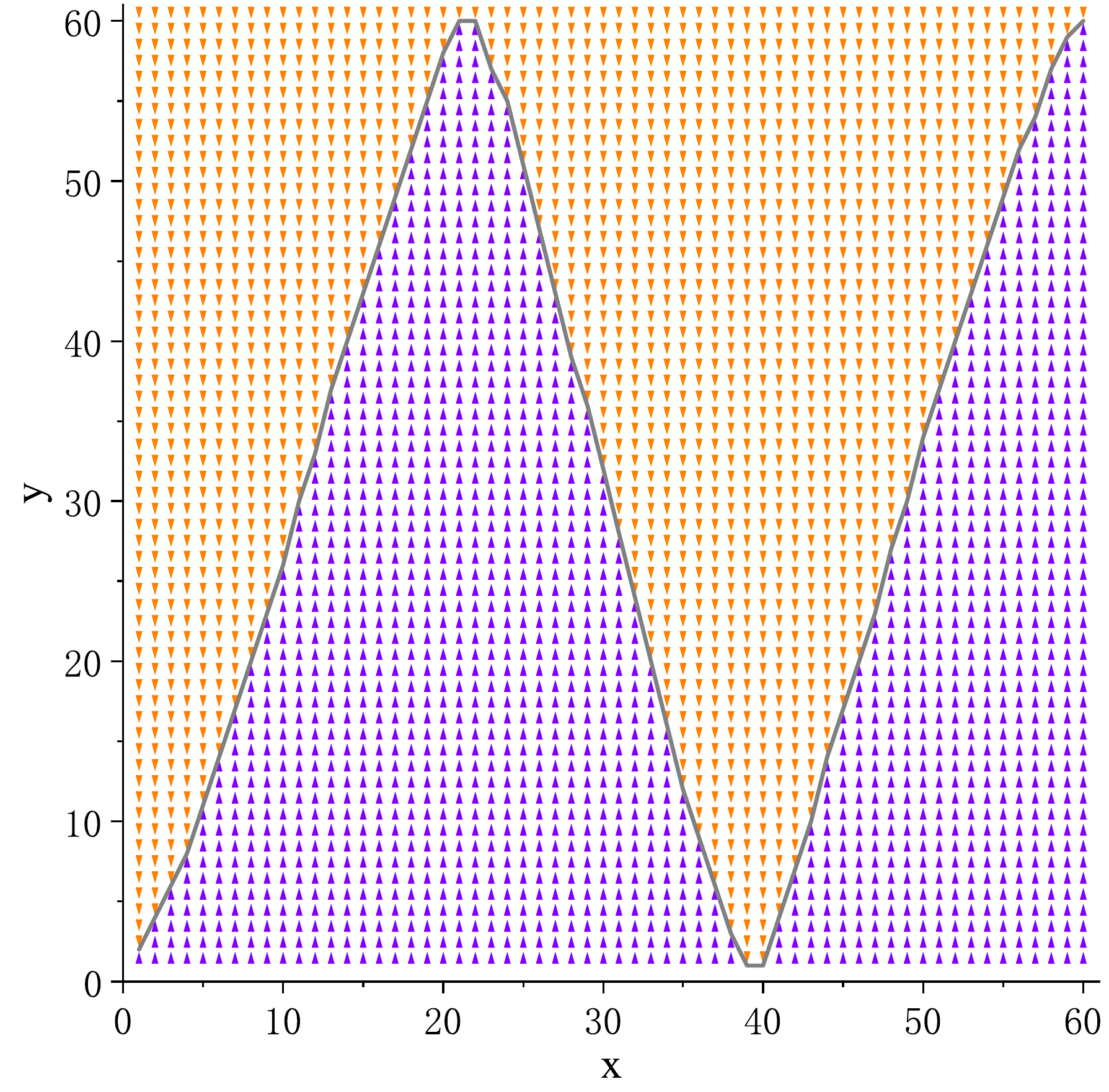}
\par\end{centering}
\caption{\label{fig:rslt-2D} sawtooth-shaped domain wall in a $60\times60$
2D lattice.}
\end{figure}
 For the 2D case, we use a $60\times60$ supercell to mimic a planar
sample. The bound change is chosen as $Z=1.16\left|e_{0}\right|$,
which is an approximate estimation from the $R3c$ phase BiFeO$_{3}$\citep{estimate_bound_change},
while the short-range interaction parameter is taken to be $J=0.00586$\,Hartree
(1 Hartree = 27.2 eV). The relative permittivity is $\varepsilon_{r}=7.164$
which renders a Coulomb energy of $Z^{2}/\varepsilon_{r}a_{0}$. We
note $J$ and $\varepsilon_{r}$ can be inferred from the parameters
used in the effective Hamiltonian for BFO\citep{Wang2012,parameters,PRB 81,PRL 99}
and the value of $J$ is compatible with the formation energy calculation
of domain walls \citep{Wang2013_domain_wall,Dieguez2013,Jiang2017}.
We note that the exact value of $J$ or $Z$ is not so important to
give rise to the sawtooth domain wall. In addition, as we will see,
the parameter $\alpha=J/\left(Z^{2}/\varepsilon_{r}a_{0}\right)$
will largely determine the configuration.

\begin{figure*}[t]
\begin{centering}
\includegraphics[width=14cm]{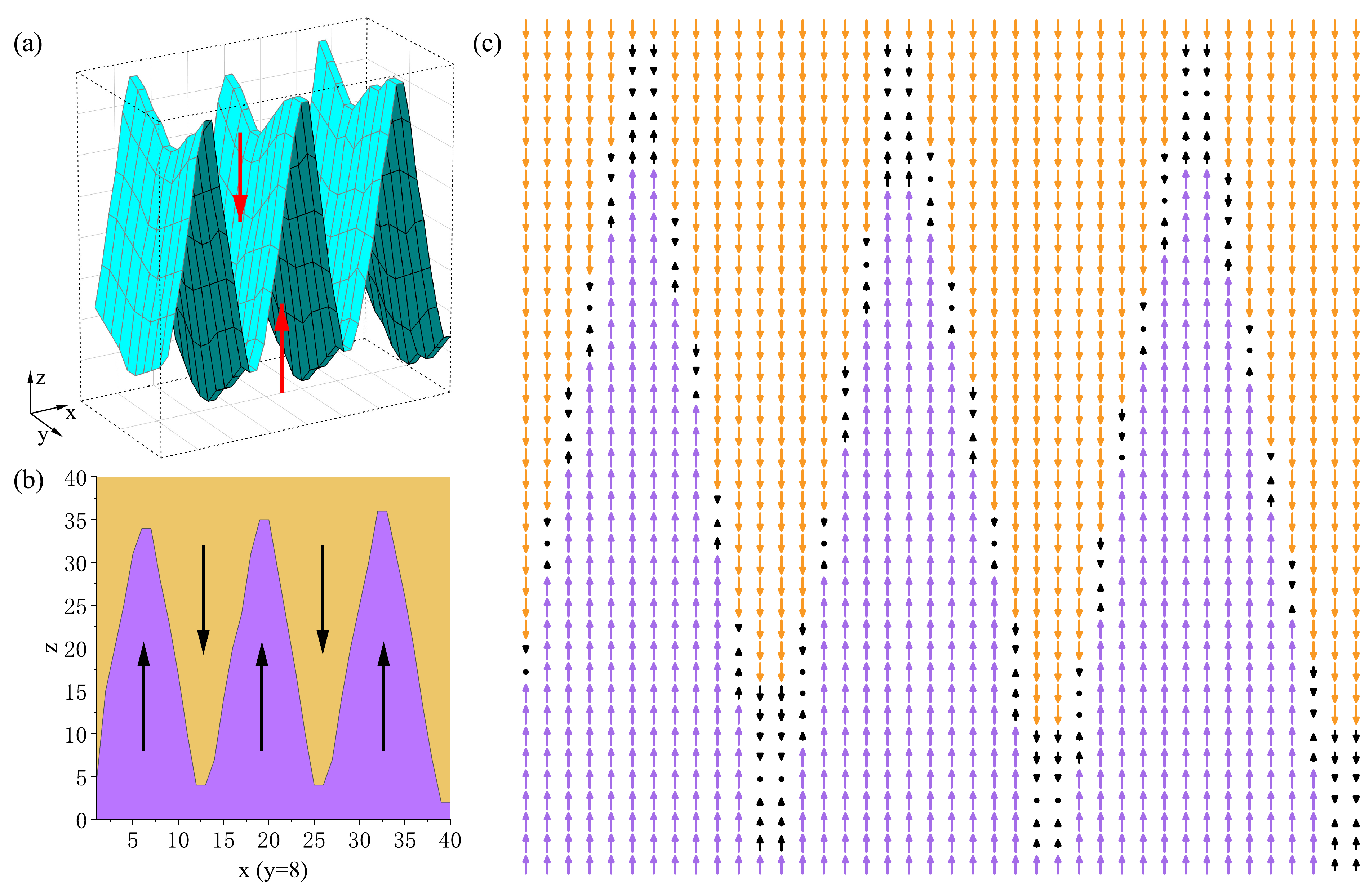}
\par\end{centering}
\caption{\label{fig:3D} Sawtooth-shaped domain wall in the 3D using a $40\times40\times10$
supercell. (a) It contains a series of conical depressions and bulges;
(b) The cross section at $y=8$ shows a jagged domain wall; (c) The
projection of the dipoles on the $x$-$z$ plane shows reduced dipoles
(black arrows) due to the average of dipoles over different $y$ sections.}
\end{figure*}
 Figure \ref{fig:rslt-2D} displays a typical 2D simulation result,
in which the sawtooth domain walls can be clearly seen. The domain
walls have an inclination of 71.47${^{\circ}}$ (the inclination will
be determined by energy analysis) and can steadily exist for 200,000
MC sweeps. For the 3D case, we use the Ewald method \citep{Wang2019},
which naturally models the periodic boundary conditions of the supercell,
to accelerate the evaluation of the Coulomb energy. The short-range
interaction is treated similarly as in 2D, except that four nearest
neighbors need to be considered instead of two. We note that, considering
experimental situation (e.g., PTO on STO where ferroelectric regions
are separated by non-ferroelectric ones), we do not assume bound charge
exist on the top-bottom boundary. Moreover, for the 2D case, direct
summation of Coulomb energy for non-periodic boundary conditions is
used for easier simulation program and avoiding complications with
the Ewald method for 2D case \citep{Santos2016}.

Using a $40\times10\times40$ supercell, we carry out 320,000 sweeps
of MC simulation at 300\,K, and the resulting domain wall is shown
in Fig. \ref{fig:3D}(a). Figure \ref{fig:3D}(b) shows the cross
section at $y=8$ where a triangular sawtooth domain wall can be clearly
seen. To compare to experimental HRTEM images, we have also projected
the dipoles along the $y$ direction, averaging along each column,
which results in Fig. \ref{fig:3D}(c). This figure not only demonstrates
the sawtooth domain walls, but can also explain the smaller dipoles
separating the two domains as observed in experiment {[}see Fig. 5(a)
of Ref. \citep{Jia2015}{]}.

\begin{figure}[h]
\begin{centering}
\includegraphics[width=8cm]{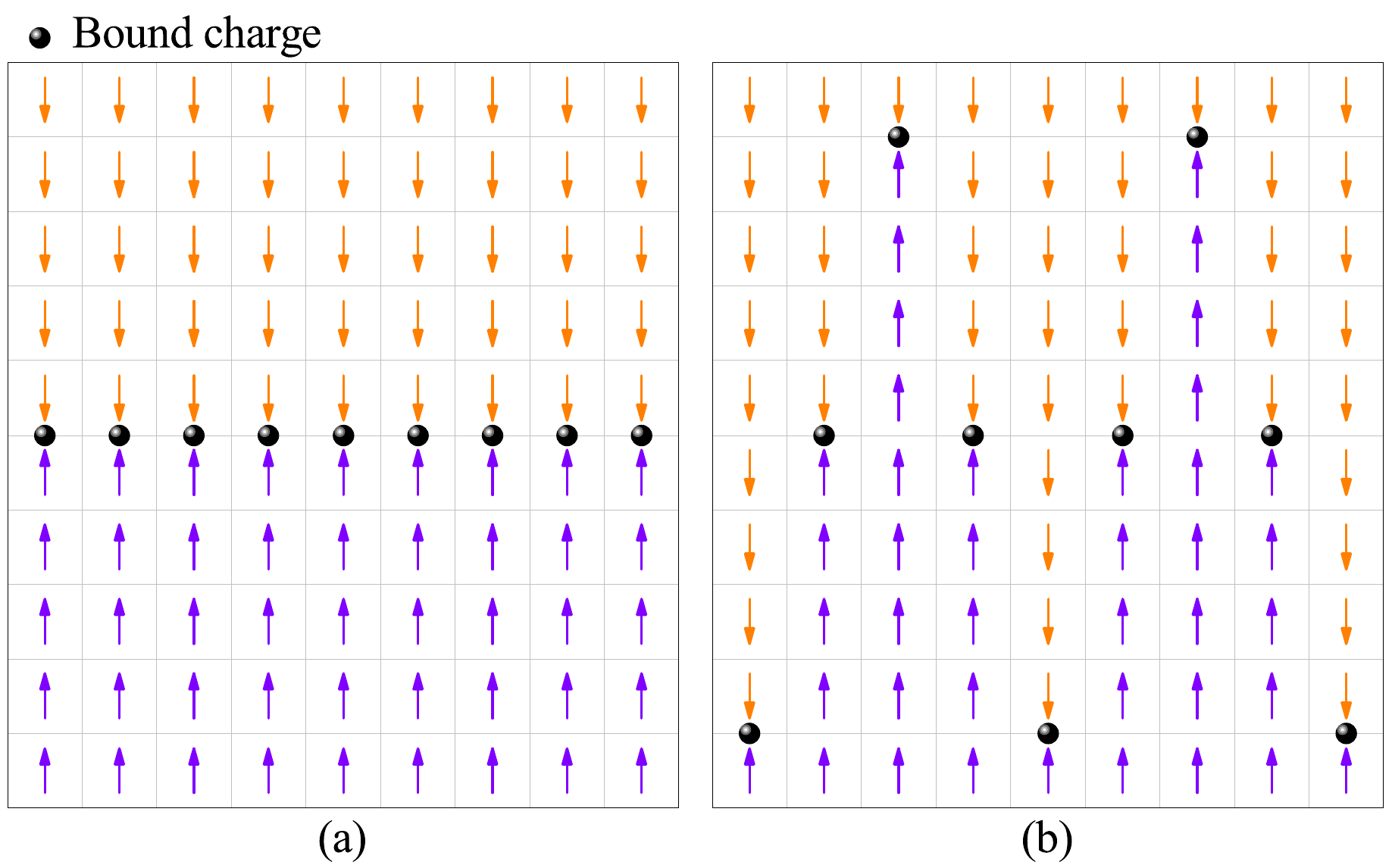}
\par\end{centering}
\caption{\label{fig:domain-change}Schematic diagram of the domain wall shape
changing with its length. (a) In this case ($l=0$, $T=\infty$),
the length of the domain wall is zero, it has the minimal short-range
interaction and maximal Coulomb interaction. (b) As the length becomes
larger ($l=32a_{0}$, $T=5a_{0}$), the domain wall appears inclined.
The Coulomb interaction decreases while the short-range interaction
increases.}
\end{figure}

\begin{figure}[h]
\begin{centering}
\includegraphics[width=8cm]{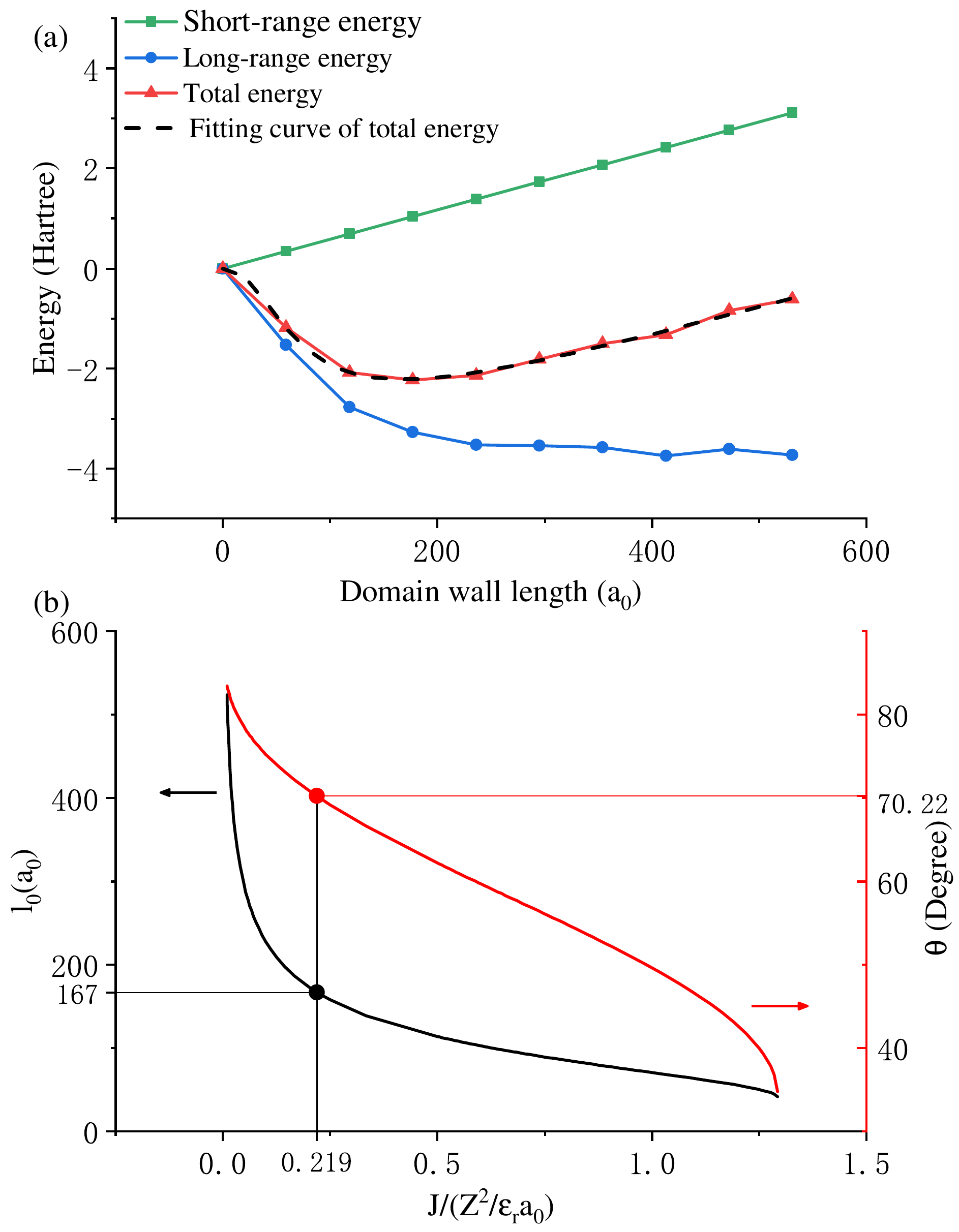}
\par\end{centering}
\caption{\label{fig:domain-wall-energy}(a) The constituent energies versus
the length of the domain wall. The short-range interaction increases
linearly, while the Coulomb interaction decreases (symbols are from
numerical computations); (b) The stabilized domain wall length ($l_{0}$,
black line) and inclination ($\theta$, red line)\textit{\emph{ versus}}\textit{
}$J/\left(Z^{2}/\varepsilon_{r}a_{0}\right)$.}
\end{figure}

As we have seen, this model, which involves only Coulomb and short-range
interactions, is adequate to reproduce the sawtooth domain walls.
With this model, it is also possible to reveal and understand how
$Z$ and $J$ can affect the domain wall morphology. To simplify the
analysis, we use the 2D case as an example and only consider triangular
sawtooth domain walls with different inclinations (see Fig. \ref{fig:domain-change}).
The length of the domain wall can be formally defined as (in unit
of $a_{0}$) $\sum_{i}\sqrt{\left(x_{i+1}-x_{i}\right)^{2}+\left(y_{i+1}-y_{i}\right)^{2}}$,
where $\boldsymbol{r}_{i}=a_{0}\left(x_{i},y_{i}\right)$ is the position
of the $i$th charge and $a_{0}$ is the lattice constant. As shown
in Fig. \ref{fig:domain-change}, $x_{i+1}-x_{i}=1$, therefore the
length is given by $\sum_{i}\sqrt{1+\left(y_{i+1}-y_{i}\right)^{2}}$.
Because the bound charge can only shift in the up and down directions,
it can be further simplified to its $y$ component as
\begin{align}
l= & \sum_{i}\left|y_{i+1}-y_{i}\right|,\label{eq:length}
\end{align}
which can unambiguously determine the triangular domain wall. One
advantage of this definition is that the short-range energy is directly
proportional to $l$ (see Fig. \ref{fig:domain-wall-energy}a), i.e.
$E^{\textrm{short}}=Jl$. The Coulomb energy also depends on $l$
as $E^{\textrm{cc}}=E^{\textrm{cc}}\left(l\right)-E_{0}^{\textrm{cc}}$
where $E_{0}^{\textrm{cc}}=Z^{2}\gamma/a_{0}$ (in unit of Hartree)
and $\gamma$ is a constant calculated according to the charge positions
shown in Fig. \ref{fig:domain-change}(a). 

As $l$ increases, the domain wall becomes sharper (i.e., the inclination
increases) {[}see Fig. \ref{fig:domain-change}(b){]}. Given a domain
wall length, we can numerically calculate its constituent energies,
which are shown as symbols in Fig. \ref{fig:domain-wall-energy}.
It can be seen that the Coulomb energy and the short-range interaction
energy show opposite trends with the length of domain wall. The short-range
interaction increases with $l$, since larger $l$ means more opposite
dipole pairs. The Coulomb energy decreases with $l$ due to the increase
of bound charge distance.

To proceed further, we propose to use $E^{\textrm{cc}}=\frac{Z^{2}\gamma}{a_{0}\varepsilon_{r}}\left(\frac{1+bl^{2}}{1+al^{2}}-1\right)$
to describe how the Coulomb energy changes with $l$, where $\gamma=220.8$
for a $60\times60$ simulation box. As a matter of fact, this expression
can be used to fit the numerically computed Coulomb energy in Fig.
\ref{fig:domain-wall-energy}, giving $a=1.99\times10^{-4}$ and $b=7.18\times10^{-5}$.
The total energy is then given by

\begin{align}
E^{\textrm{tot}}= & Jl+\frac{Z^{2}\gamma}{\varepsilon_{r}a_{0}}\left(\frac{1+bl^{2}}{1+al^{2}}-1\right),
\end{align}
which can be derived by first considering the decrease of the long-range
energy with $l$ (the blue line with solid circles in Fig. \ref{fig:domain-wall-energy}(a)).
The variation of the total energy with $l$ is also shown in Fig.
\ref{fig:domain-wall-energy}(a). The equilibrium domain wall length
$l_{0}$ can be found by minimizing the total energy with respect
to $l$, and its dependence on $J/(Z^{2}/\varepsilon_{r}a_{0})$ is
shown in Fig. \ref{fig:domain-wall-energy}(b). This result indicates
that the parameter $\alpha\equiv J/(Z^{2}/\varepsilon_{r}a_{0})$
is crucial for determining the domain morphology (which is independent
of their absolute value) and larger $\text{\ensuremath{\alpha}}$
tends to bind the bound charges closer to each other.

In the simulations that generate Figs. \ref{fig:rslt-2D} and \ref{fig:3D},
$\alpha=0.22$ is used. The resulting domain wall length and sawtooth
period are consistent with the theoretical estimation. In numerically
obtaining $l_{0}$ for Fig. \ref{fig:domain-wall-energy}(b), we find
that when $\alpha>1.29$, no solution can be found for $l_{0}$, which
is consistent with our numerical findings (not shown here)\textbf{
}that arbitrarily chosen $J$ and $Z$ cannot support the existence
of such domain walls.

It shall be noted that the precondition for the above analysis is
that triangular domain walls already exist. The constraint of $\alpha<1.29$
can be understood by estimating the two energies of the configuration
shown in Fig. \ref{fig:domain-change}(b). Assuming that two neighboring
bound charges are shifted by $y$ vertically, the short-range interaction
is $NJy$ ($N=60$ for the $60\times60$ simulation box), while the
Coulomb energy pertaining to this configuration is the horizontal
line of bound charges {[}Fig. \ref{fig:domain-change}(a){]} tilted
by an angle of $\theta$ ($\tan\theta=y$), giving the energy of $\left(Z^{2}\gamma/\varepsilon_{r}a_{0}\right)$$\left(1/\sqrt{1+y^{2}}-1\right)$.
Since the Coulomb energy and the short-range interaction energy shall
balance each other (not that one overwhelms the other) and reduce
the total energy, therefore $NJy+Z^{2}\gamma/\varepsilon_{r}a_{0}\left(1/\sqrt{1+y^{2}}-1\right)<0$
is necessary, resulting in $\alpha<\gamma\left(1-1/\sqrt{1+y^{2}}\right)/\left(Ny\right)\leq1.10$
for $N=60$, where the maximum is reached when $y=1.27$. Since the
Coulomb energy in the triangular case shall be larger than this value
as the bound charges are closer, the final value of $\alpha$ shall
be smaller than 1.10.

In fact, a more stringent constraint can be obtained with Fig. \ref{fig:domain-change}(a)
as the initial configuration and consider only one bound charge (the
first one from the left) is shifted upward by $y$, which satisfies
$Jy+\left(Z^{2}/\varepsilon_{r}a_{0}\right)\sum_{n=1}^{N-1}\left(1/\sqrt{n^{2}+y^{2}}-1/n\right)<0$
or $\alpha<\sum_{n=1}^{N-1}\left(1/n-1/\sqrt{n^{2}+y^{2}}\right)/y\leq0.42$
where the maximum is reached when $y=1.7$. This result\textbf{ }further
constrains the parameters that can form sawtooth domain walls, indicating
that there is an upper bound for $\alpha$ to make the sawtooth domain
walls possible. This constraint, which is necessary to form sawtooth
domain walls, is also verified using MC simulation. For the 3D case,
using the parameters chosen for BFO, we found when $\alpha<0.806$,
the sawtooth domain wall is possible.

In addition to the constraint on $\alpha$, it was pointed out that
180${^{\circ}}$ domain occurs when no (or very small) epitaxial strain
are applied from the substrate, while 90$^{\circ}$ or other domain
patterns are expected with larger values \citep{chinese_sawtooth,AM50(2002),Ma}.
This can be understood with the strain-dipole coupling \citep{Zhong},
where the dipoles experience extra energy from strain, which likely
makes their flipping more difficult comparing to a partial rotation
of 90$^{\circ}$, effectively increasing the $J$ parameter. In addition,
similar to magnetic domain walls \citep{Kasap}, dislocation, impurity,
and defect can also hinder domain wall growth.

The effect of these parameters on the sawtooth domain walls will be
further discussed later. Since charged domain walls can be compensated
by free charge carriers \citep{charged-domain-walls,charged-domain-walls2},
the tendency of the wall to be inclined is reduced as the Coulomb
interaction is reduced ($\alpha$ becomes larger). Depending on the
value of $J$, the sawtooth domain could still exist if $J$ is small
enough. On the other hand, we need to note that charged domain walls
are not always compensated by free carrier, such as the hybrid perovskite
materials \citep{Chen2018}.

In summary, we have built a minimal model to reveal the origin of
the sawtooth-shaped domain walls observed in ferroelectric materials.
Our model based MC simulations show that the competition between the
long-range Coulomb energy from bound charges and the short-range interaction
energy are responsible for the formation of these peculiar domain
walls. Further analysis also shows that the combined parameter $J/\left(Z^{2}/\varepsilon_{r}a_{0}\right)$
is critical in determining the inclination of the sawtooth-shaped
domain walls and its value has to satisfy certain conditions for this
unique type of domain walls to appear in ferrolectrics.\\

\begin{acknowledgments}
This work is financially supported by the National Natural Science
Foundation of China (Grant No. 11574246, U1537210, 51671194, and 11974268),
National Basic Research Program of China (Grant No. 2015CB654903,
2014CB921002), and the Key Research Program of Frontier Sciences CAS
(QYZDJ-SSW-JSC010). D.W. also thanks the support from China Scholarship
Council (201706285020). L.B. acknowledges ARO Grant No. W911NF-16-1-0227.
\end{acknowledgments}


\begin{thebibliography}{99}
\bibitem{parameters}It shall be note that $\varepsilon_{r}$ is directly
taken from the effective Hamiltonian \citep{Zhong}. One the other
hand, in the effective Hamiltonian, the short-range interactions are
kept to the 3rd nearest neighbors described by seven parameters ($j_{1}$
to $j_{7}$, see Fig. 1 of Ref. \citep{Zhong}). Here in the minimal
model, their effects are summarized in a single parameter $J$, which
should not be directly compared to the values used obtained from the
formation energy.
\end{thebibliography}
\end{document}


\title{Supplemental Material}
\author{J. Zhang}
\affiliation{School of Microelectronics \& State Key Laboratory for Mechanical
Behavior of Materials, Xi'an Jiaotong University, Xi'an 710049, China}
\author{Y.-J. Wang}
\affiliation{Shenyang National Laboratory for Materials Science, Institute of Metal
Research, Chinese Academy of Sciences, Wenhua Road 72, 110016 Shenyang,
China}
\author{J. Liu}
\affiliation{State Key Laboratory for Mechanical Behavior of Materials, School
of Materials Science and Engineering, Xi'an Jiaotong University, Xi'an
710049, China}
\author{J. Xu}
\affiliation{School of Microelectronics \& State Key Laboratory for Mechanical
Behavior of Materials, Xi'an Jiaotong University, Xi'an 710049, China}
\author{D. Wang}
\email{dawei.wang@xjtu.edu.cn}

\affiliation{School of Microelectronics \& State Key Laboratory for Mechanical
Behavior of Materials, Xi'an Jiaotong University, Xi'an 710049, China}
\author{L. Wang}
\affiliation{Electronic Materials Research Laboratory, School of the Electronic
and Information Engineering, Xi'an Jiaotong University, Xi'an 710049,
China}
\author{X.-L. Ma}
\affiliation{Shenyang National Laboratory for Materials Science, Institute of Metal
Research, Chinese Academy of Sciences, Wenhua Road 72, 110016 Shenyang,
China}
\author{C.-L. Jia}
\affiliation{School of Microelectronics \& State Key Laboratory for Mechanical
Behavior of Materials, Xi'an Jiaotong University, Xi'an 710049, China}
\affiliation{\textsuperscript{}Ernst Ruska Center for Microscopy and Spectroscopy
with Electrons, Research Center Jülich, D-52425 Jülich, Germany}
\author{L. Bellaiche}
\affiliation{Department of Physics and Institute for Nanoscience and Engineering,
University of Arkansas Fayetteville, Arkansas 72701, USA}
\date{\today}

\maketitle
In this supplemental material we show that adding another factor to
the minimal model will make our results compare even better to experiments.
This further confirms that our model indeed captures the most important
factors that give rise to the sawtooth domain walls.

\begin{figure}[h]
\begin{centering}
\includegraphics[height=10cm]{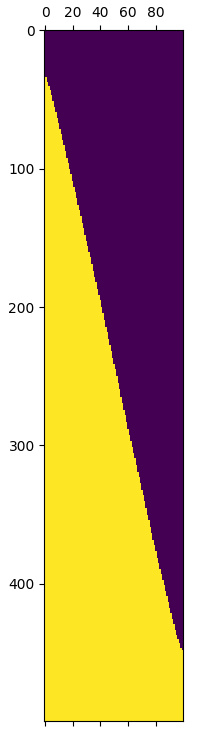}
\par\end{centering}
\caption{Inclination flat domain wall can arise with the minimal model when
the thickness is much larger than the width of the simulation box.\label{fig:Inclination-flat-domain}}
\end{figure}
 We have shown that the the minimal model shows that inclined domain
walls can develop in ferroelectrics, which can turn into sawtooth
domain walls when they encounter interfaces or interacting with other
domain walls. The reason is that the inclined flat domain wall will
eventually contact with the interface resulting in effective charges
accumulating along a straight line, which has higher energy than the
swatooth domain wall. However, in the extreme case that the thickness
is much larger than its width (which is not very likely ), the inclined
flat domain wall can indeed be seen in the 2D simulations as seen
in Fig. \ref{fig:Inclination-flat-domain}.

To address this issue, other factors need to be considered and added
on top of those adopted in the minimal model. The most likely factor,
in our opinion, is that the Coulomb interaction range is limited due
to screening imposed by the dipoles in the background. In fact, Zhou
\emph{et al} indicate that the locking of the inclined flat domain
wall may contribute to the formation of the zig-zag domain wall \citep{Zou2018},
which hints toward the fact that far-away domain walls develop independently
of each other and become zig-zag when they eventually meet and interlock.
This observation supports the idea to use a Coulomb interaction of
limited range.

As a matter of fact, while the minimal model has absorbed dipoles
to the background, they still exist in the real system and reorient
slightly in accordance with the electric field they experience. Such
dipole dynamics, while small, could effectively screen the interaction
between bound charges, resulting in short-range Coulomb interaction
(e.g., like the Yukawa potential). Furthermore, bound charges can
attract free charge carriers or charge defects, which can compensate
the bound charge, effectively reducing the interaction between the
bound charges \citep{Rojac2017}. To address such effects, we can
add additional a factor to our model, i.e., using a finite Coulomb
interaction range.

\begin{figure}[h]
\begin{centering}
\includegraphics[width=10cm]{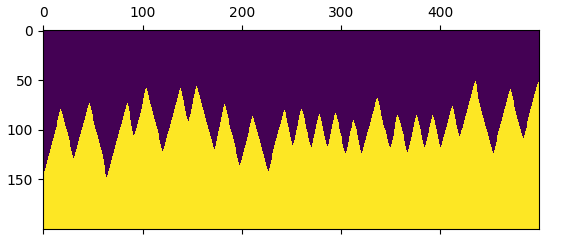}
\par\end{centering}
\caption{\label{fig:Simulation-results-after}Simulation results for the 2D
case when Coulomb interaction range is limited.}
\end{figure}
 Figure \ref{fig:Simulation-results-after} the sawtooth domain walls
in 2D after limiting Coulomb interaction range. In this simulation,
we have used a $500\times200$ simulation box, and carry out 320,000
sweeps of MC simulation at 300\,K. Here we still use the parameters
in the manuscript for the simulation. The short-range interaction
parameter is taken to be $J=0.00586$\, Hartree. The effective bound
change $Z=1.16\left|e_{0}\right|$ and the relative permittivity $\varepsilon_{r}=7.164$
are used in this simulation, similar to those used in the main text.
However, in this simulation, we have limit the Coulomb between two
bound charges to 15$a_{0}$ With such a model, The simulation results
clearly show that the sawtooth domain wall can exist stably in the
system, not affected by the size of the simulation box. However, we
have verified that the domain wall morphology, and the period in particular,
is affected by the interaction strength and the range of the Coulomb
interaction specified.

\begin{figure}[h]
\begin{centering}
\includegraphics[width=12cm]{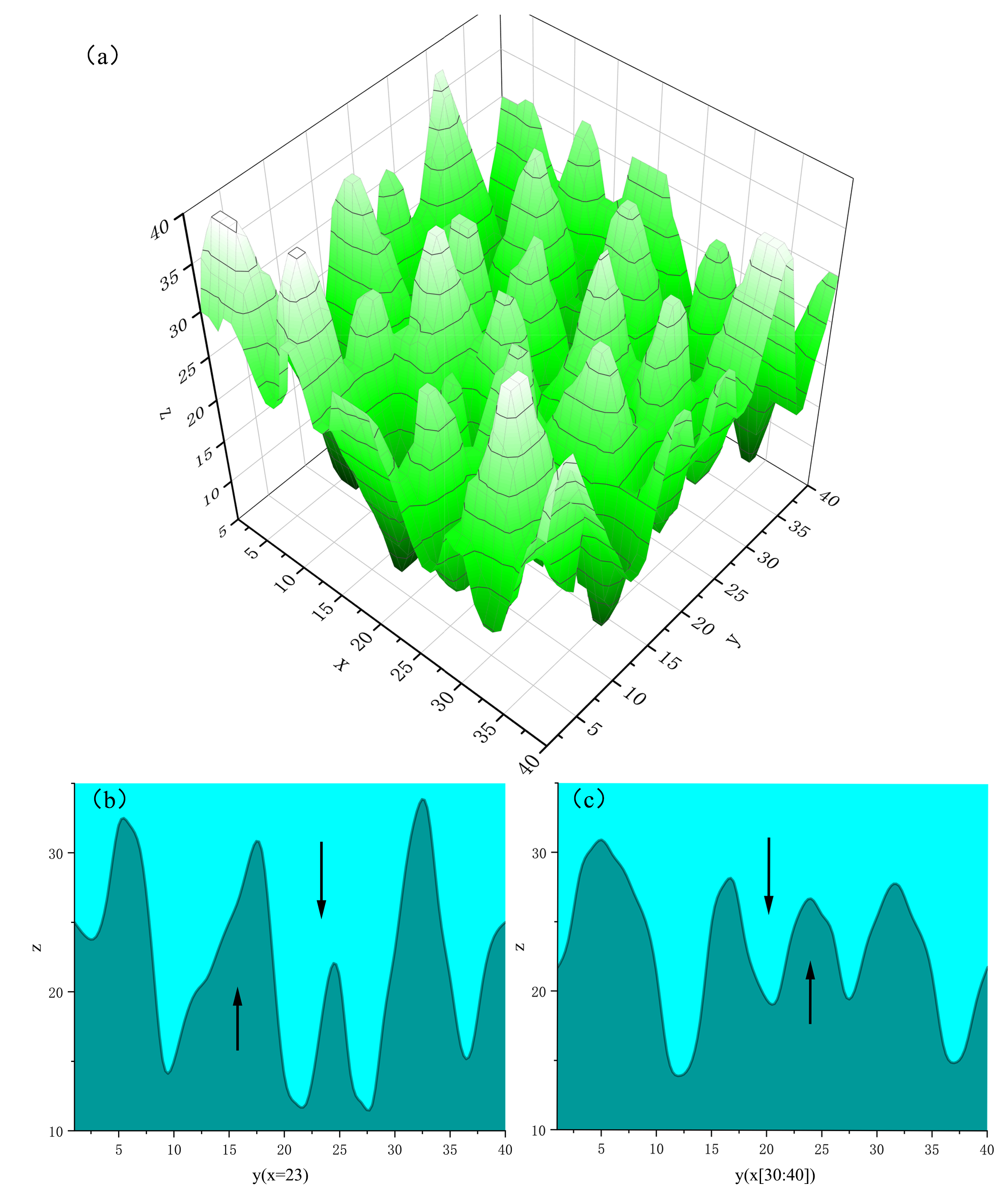}
\par\end{centering}
\caption{\label{fig:3D result}Simulation results for the 3D case when the
Coulomb interaction range is limited. (a) 3D view of the simulation
results; (b) A cross section in the $y$-$z$ plane obtained at $x=23$;
(c) A projection of 10 $y$-$z$ layers along the $x$ direction ($x=30\dots40$).}
\end{figure}
 For the 3D case, we empoly the Ewald method and periodic boundary
conditions to perform the simulation in a $40\times40\times40$ box
with the same parameters as in the 2D case. In addition, we also set
the limitation on the Coulomb interaction range to 10$a_{0}$ (since
the simulation box size is smaller than the 2D case). After 30,000
Monte Carlo sweeps (i.e., $30000\times40^{2}$ attempts to move the
domain wall), the simulation results show domain walls converges to
a series of conical peaks as shown in Fig. \ref{fig:3D result}(a).
Figure \ref{fig:3D result}(b) shows a cross section of the domain
wall in the $y$-$z$ plane at $x=23$ and Fig. \ref{fig:3D result}(c)
shows a cross section o shows a projection of 10 $y$-$z$ layers
along the $x$ direction (to compare to the results shown in the main
text) where the position of the domain wall is determined by averaging
the positions in those 10 layers. Both of Fig. \ref{fig:3D result}(b)
and (c) clearly show sawtooth domain walls.

\begin{figure}[h]
\begin{centering}
\includegraphics[width=10cm]{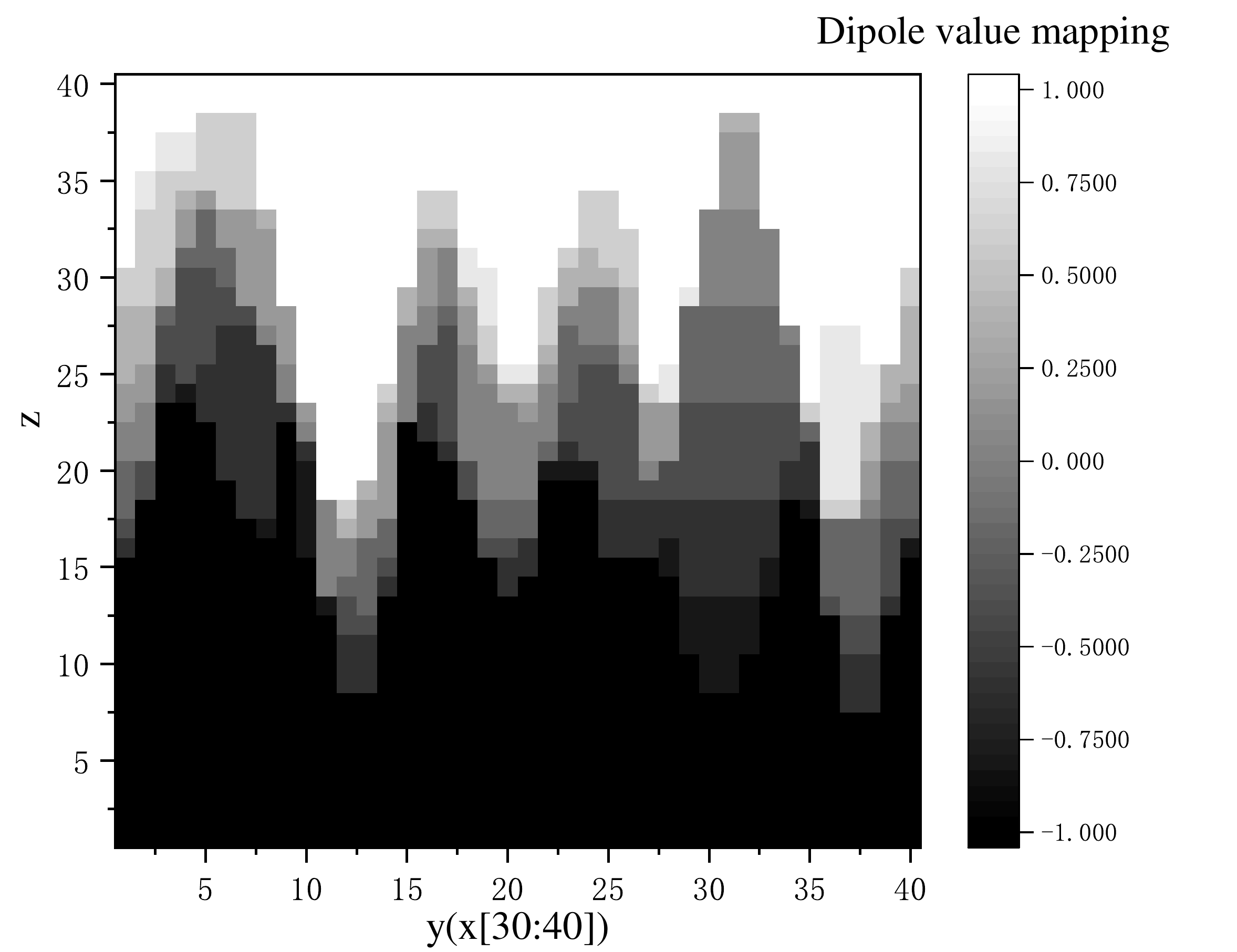}
\par\end{centering}
\caption{\label{fig:Dipole-intensity-projection} Dipole magnitude map for
the $y$-$z$ layers ($x=30\dots40$) projected along the $x$ direction
where the dipoles are averaged over all the 10 layers.}
\end{figure}

An important feature for the 3D case is that, whether we cut the sample
along $x$ or $y$, the sawtooth wall will always appear. As a matter
of fact, Jia \emph{et al} have indicated that the striped domains
are filled with sawtooth domain walls that can be seen along different
projection directions \citep{Jia2015}, supporting our simulation
results. In addition, Fig. \ref{fig:Dipole-intensity-projection}
shows a similar projection as Fig. \ref{fig:3D result}(c), but also
records the averaged dipole values with negative (positive) values
indicate dipoles pointing upward (downward), which may be used to
explain the smaller dipoles seen on the domain wall as seen in experiments
\citep{Jia2015}. Finally, we have also tried simulation boxes with
different sizes, obtaining qualitatively similar results.

